\documentclass{iau}

\usepackage{amsmath}
\usepackage{graphicx}
\usepackage{multirow}
\usepackage{soul}
\usepackage{subcaption} 
\setulcolor{red}

\def\apj{{ApJ}}

\def\apjl{{ApJ Letters}} 
\def\apj{{ApJ}} 
\def\aap{{A\&A}}   
\def\mnras{{MNRAS}}

\def\apjs{{ApJ Supplement}}

\newcommand{\solphys}{{\it Solar Phys.}}

\newcommand{\Fig}[1]{Figure~\ref{#1}}
\newcommand{\Tab}[1]{Table~\ref{#1}}
\newcommand{\Sec}[1]{Section~\ref{#1}}
\newcommand{\bmax}{$B_{\rm max}$}
\newcommand{\bmean}{$B_{\rm mean}$}

\begin{document}

\doival{ }

\aopheadtitle{Proceedings IAU Symposium No.365, 2023}
\editors{A. Getling \&  L. Kitchatinov, eds.}

\title{AutoTAB: An Algorithm to Track Bipolar Magnetic Regions and Initial results from the Tracked BMRs}

\author{Anu Sreedevi$^1$, Bibhuti Kumar Jha$^2$, Bidya Binay Karak$^1$ and\\ Dipankar Banerjee$^3$}
\affiliation{$1$ Department of Physics, Indian Institute of Technology (Banaras Hindu University), Varanasi 221005, India\\ email:\email{anubsreedevi.rs.phy20@itbhu.ac.in}}
\affiliation{$2$ Southwest Research Institute, Boulder 80302, USA}
\affiliation{$^3$Aryabhatta Research Institute of Observational Sciences, Nainital 263002, Uttarakhand, India}


\begin{abstract}
AutoTAB is a state-of-the-art, fully automatic algorithm that tracks the  Bipolar Magnetic Regions (BMRs) in magnetogram observations. AutoTAB employs identified BMR regions from Line-of-Sight magnetograms from MDI and HMI (1996--2022) to track the BMRs through their evolution on the nearside of the Sun. AutoTAB enables us to create a comprehensive and unique catalog of tracked information of 9232 BMRs in the mentioned time period. This dataset is used to study the collective statistical properties of BMRs 
and particularly to identify the correct theory for the BMR formation.
Here, we discuss the algorithm's functionality and the initial findings obtained from the AutoTAB BMRs catalog.
\end{abstract}


\maketitle

\section{Introduction}
Bipolar Magnetic Regions (BMRs) are the visible proxies of magnetic solar activity. The sunspot pairs observed in white-light images of the Sun translate to opposite polarities of BMRs as seen in Line-of-Sight (LOS) magnetic field observations. The number of BMRs appearing on the solar disk shows a prominent cyclic behavior in accordance with the 11-year solar cycle. These regions are observed to emerge in a tilted fashion with respect to solar E-W direction \citep{HE1919}. Evolution and dispersal of such regions on the surface, along with the meridional flow, causes the reversal in the poloidal field of the Sun 
\citep{B1961, L1964, Kar20, MK2022, EB2023}. Consequently, the investigation of BMRs, their origin, and their evolutionary dynamics has been a compelling area of research in the past from both theoretical and observational perspectives \citep{KM17, KM18, BKK23}. 

A regular magnetic field observation of the Sun has been performed since the early 1970s using ground-based solar observatories. In the last few decades, with the advancement of space-based observatories/telescope, such as Michelson Doppler Image \citep[MDI;][]{SB1995} and Helioseismic and Magnetic Imager \citep[HMI;][]{SS2012} we have been provided with the continuous full disk LOS magnetograms from 1996 onwards. Since then, magnetogram data has been used to monitor BMRs constantly to study solar magnetism and to forecast solar eruptions. Diverse methods, including automated techniques, have been developed for this purpose \citep{SK2012, TP2014, JW2016}. However, a consistent, calibrated catalog of tracked BMR information for solar magnetism study is still absent. To address this, we introduce AutoTAB, an advanced automatic tracking algorithm for BMRs \citep{SJ2023Symp, SJ2023}. In \Sec{sec:algorithm}, we explain the details and workings of the algorithm.

The theory behind the formation of BMRs is still debated. According to the thin-flux tube model, BMRs are believed to be formed when magnetically buoyant flux tubes from the convection zone rise up to the solar surface \citep{P1955}. The diverging flows at the top of the rising flux tubes will be subjected to Coriolis force to form the tilt in the BMRs \citep{DC1993}. However, backing for this study remains limited. Here, we utilize tracked BMR data to assess the thin-flux tube model's validity in explaining BMR emergence. 
Many prior studies have explored this using various sunspot catalogs and before-mentioned automated methods \citep{WS1991, SK2012, JK2020}. However, some of these studies treat each BMR detection as a new BMR. This approach may introduce bias in analysis, as larger, longer-lasting BMRs contribute to a higher weightage. Our study addresses this issue by actively tracking the BMRs.

\section{Data and Method} \label{sec:algorithm}

In this study, we have used the full disk LOS magnetogram data with the cadence of 96 minutes for the period of September 1996–December 2022 (Cycles 23
and 24) from MDI \citep[1996--2011;][]{SB1995} and 
HMI \citep[2010--present;][]{SS2012} onboard the Solar and Heliospheric Observatory(SOHO) and Solar Dynamic Observatory (SDO), respectively. The BMRs are automatically identified using the method outlined in \citet{SK2012}, as adapted with slight modification by \citet{JK2020}. The identified BMRs are then saved as rectangular regions in the form of binary files. These rectangular regions also include extra pixels that may not be part of the BMR, which can potentially affect tracking efficiency, particularly in the case of closely lying BMRs. To isolate BMR boundaries more accurately, a threshold of 100~G is used, followed by the morphological operations (using \textit{morp\_close} function in IDL) with circular kernels of different sizes on these pixels. Once the boundaries of BMR are identified precisely, we extract morphological, magnetic, and location information for each identified region.

AutoTAB utilizes binary masks of identified BMRs to track them during their appearance on the near side of the Sun. The method of tracking is based on feature association similar to the one used for sunspot tracking in \citet{JP2021}. A brief description of the tracking algorithm follows: A BMR intended to be tracked is selected and isolated to a separate binary file (BF-I) and assigned with a unique AutoTAB ID. From the positional information of the BMR, we calculate the tracking period (T$_{\rm max}$), i.e., the time taken for the BMR to reach the solar West limb. This indicates the duration during which AutoTAB will scan for isolated BMR. Next, we gather all binary masks within the time period T$_{\rm max}$ (BF-1, BF-2, etc.). Then, BF-I is differentially rotated with respect to all the collected binary masks (BF-1, BF-2, etc.) to evaluate the overlap between each of them. If the overlap is found to be greater than 150 pixels, those regions are identified with the same AutoTAB-ID. If the overlap condition is not satisfied for the next two-day period of time, the overlap criterion is modified to 250 pixels to ensure that the BMR tracked is the same one. A detailed description of the working of the tracking algorithm and details of the pre-processing step is provided in \citet{SJ2023}.

\begin{figure}
  \centering
  \begin{subfigure}{\textwidth}
    \includegraphics[width=\textwidth]{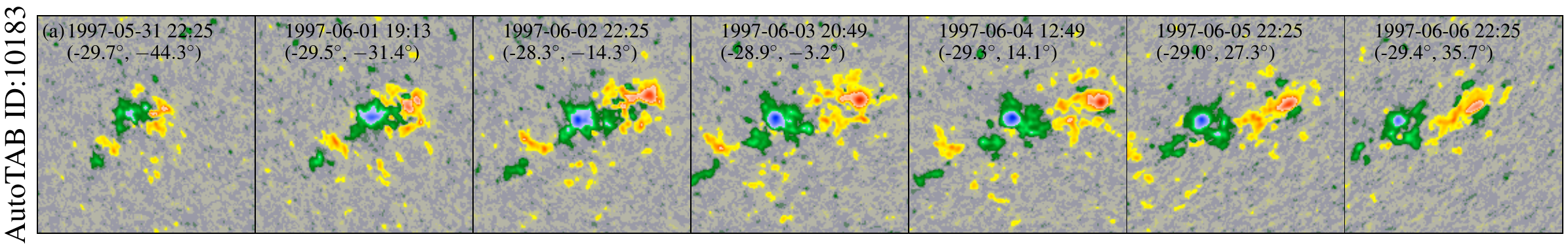}
  \end{subfigure}
  \begin{subfigure}{\textwidth}
    \includegraphics[width=\textwidth]{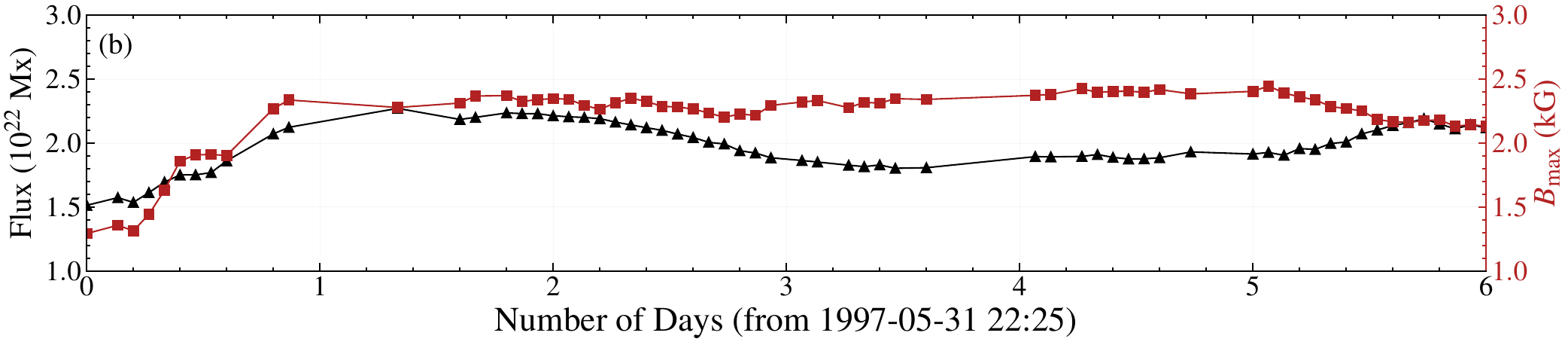}
  \end{subfigure}
  \caption{Panels (a) show the snapshots of the tracked BMR, AR08048 (AutoTAB-ID 10183), corresponding to each day during the tracking. Panels (b) show the evolution of total flux and \bmax\ for the tracked BMR.}
  \label{fig:fig1}
\end{figure}

\Fig{fig:fig1}(a) is a representative example of the tracking of AutoTAB. It shows the evolution of AR08048 (AutoTAB ID 10183) observed in MDI. This BMR was first identified near the East limb ($29.7^{\circ}$~S, $44.3^{\circ}$~E), and AutoTAB tracks it for the next seven days (in 65 observations) till it starts decaying close to the West limb. The variation of total flux and \bmax is shown in  \Fig{fig:fig1}(b) over the tracking period. 

\section{Results}
AutoTAB has successfully identified and tracked 9232 BMRs in Cycles 23 and 24 for the time period between 1996--2022. On a closer look at the tracked BMRs, we find that BMRs were tracked for a broad range of lifetimes, from hours to weeks, and over large flux ranges throughout their journey on the nearside of the Sun. 
BMR lifetime is defined as the duration for which AutoTAB maintains tracking. In instances of poor tracking likely resulting from data corruption or absence, those BMRs were excluded from the dataset. The remaining BMRs were categorized into three distinct classes.
\begin{itemize}
    \item Short-lived (SL): These are the BMRs that have a lifetime of less than 8 hours in the nearside of the Sun. They are generally small BMRs with small areas within the flux range in order of $10^{19}$ or $10^{20}$~Mx. Mostly, they are the ephemeral region, and a detailed study will be done for this class in the follow-up manuscript.
    \item Diskpassage (DP): This class accommodates such BMRs that AutoTAB has tracked in its evolutionary stages. That includes the BMRs that emerge in the nearside of the sun but continue evolving in the farside beyond 45$^{\circ}$W, the BMRs that emerge on the far side and decay in the nearside between  45$^{\circ}$E-W, and finally, the BMRs that emerge and decay on the far side. This class contains bigger BMRs within the flux range of $10^{22}$~Mx.
    \item Lifetime (LT): Any BMRs that emerge and decay in the nearside, between 45$^{\circ}$E-W, constitutes this class. The lifespan of BMRs from this class is close to a week.
\end{itemize}

\subsection{Statistical Behaviour of Tracked BMRs}
\begin{figure}[t]
\centering
\includegraphics[width=\textwidth]{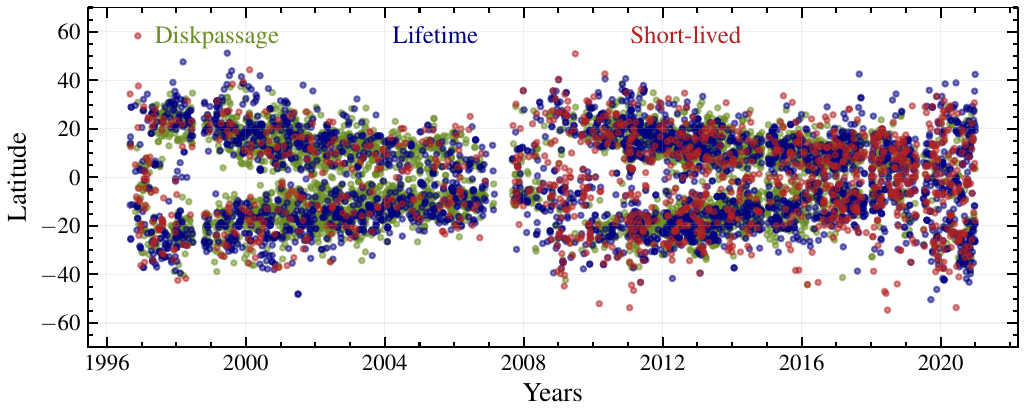}
\caption{Latitude–time distribution (butterﬂy diagram) of the tracked BMRs for (a) SL (red), (b) LT (blue), and (c) DP (green) classes.}
\label{fig:fig2}
\end{figure}

To explore the statistical behavior of the tracked BMRs, we explore if the emergence of BMRs obeys the expected solar cycle behavior by looking at well-known latitudinal time distribution in the form of a butterfly diagram. \Fig{fig:fig2} represents the butterfly diagram, and each point represents each individually tracked BMR and the latitude of each BMR is determined at its peak flux during tracking, and the same applies to other properties (\bmax, \bmean, tilt). DP and LT class BMRs conform to the butterfly wing pattern across the tracked years. A definite wing-like pattern 
is not evident in the SL class as the points are scattered over all the latitudes at all times, but we see their latitudes limited between $\pm60^{\circ}$, suggesting that they are the byproducts of large-scale toroidal magnetic ﬁeld \citep{KB2016}. The monthly new emergence of BMRs in each class exhibits a similar pattern. LT and DP class BMRs display consistent solar cycle variations, aligning with fluctuations in the number of new emergences. In the case of the SL class, we could not observe such behavior, and a sufficient number of such BMRs were detected even during solar minima when the toroidal field of the sun is at a minimum (See Figure 6 in \citet{SJ2023}). The number of total BMRs tracked in each class, along with their average magnetic properties, are listed in \Tab{tab:tab1}. A more detailed description of the mentioned classes and their statistics is explored in \citet{SJ2023}.

\begin{table}
    \centering
    \begin{tabular}{|c|c|c|c|c|}
         \hline
         Classiﬁcation  & $\#$ of BMRs &  Flux $\pm\Delta$Flux  & \bmax $\pm\Delta$\bmax & \bmean $\pm\Delta$\bmean\\
         &  & ($10^{22}$ Mx)  & (G) & (G)\\
         \hline
         Short-lived (SL)  & 1251   & 0.26 $\pm$ 0.01  & 541.32 $\pm$ 5.38 & 197.46 $\pm$ 0.66\\ 
         Lifetime (LT)     & 3191   & 1.50 $\pm$ 0.02  & 949.20 $\pm$ 6.35 & 224.46 $\pm$ 0.81\\
         Disk Passage (DP) & 4710   & 2.05 $\pm$ 0.01  & 1436.83 $\pm$ 7.02 & 281.07 $\pm$ 0.71\\
         \hline
    \end{tabular}
    \hspace{0.5cm}
    \caption{Some key parameters of different classes of tracked BMRs.}
    \label{tab:tab1}
\end{table}

\subsection{Viability of thin-flux tube model for explaining BMR emergence}
In this section, we delve into an examination of whether the tracked information from AutoTAB aligns with the expected outcomes proposed by the thin-flux tube model. The  thin-flux tube simulations by \citet{FF1994} proposed the following  empirical relation.

\begin{equation}
\gamma \propto \mathrm{sin}\lambda B_0^{-5/4} \Phi^{1/4},
\end{equation}
where $B_{0}$ is the initial magnetic field strength in the flux tube, $\Phi$ the magnetic flux inside the rising flux tubes.
From this theory, it is expected that higher the initial magnetic field of the tube, the more quickly flux tubes will rise to the surface, and the Coriolis force will have less time to produce the tilt in BMRs. Furthermore, as the flux in the loop increases, tilt will also increase due to the effect of drag force. A point to note here is that, in the simulation studies, flux in the $\Omega$-loop is independent of the 
initial magnetic field in the flux tube. 
From the tracked BMR data, we 
 indeed
 find a good correlation between the measured unsigned flux and \bmean.
 Also 
in the observations, the magnetic field values measured for the BMRs may not be 
the same as the 
 initial magnetic field of the BMR forming flux tube. 
Considering 
these,
the following are some of the observational evidence we can expect from the thin-flux tube model: 
\begin{itemize}
\item The footpoint separation to increase over the BMR's lifetime. The footpoint separation here is defined as the great circle distance between the two polarities of the BMR. If the flux tubes are rising, we should expect that the distance between the polarities should increase as the BMR matures and gets saturated towards the end of life.
\item If Coriolis force is the reason for tilted BMRs, then BMRs should emerge with tilt, and we can expect well-defined Joy's Law even from the initial phase of tracking. 
\item We can also assume that the measured unsigned flux within these regions will likely influence their tilt characteristics.
\end{itemize}

In the rest of the section, we assess each of the mentioned observations based on the findings from AutoTAB within the framework of the thin-flux tube model.\\

\begin{figure}[t]
\centering
\includegraphics[width=\textwidth]{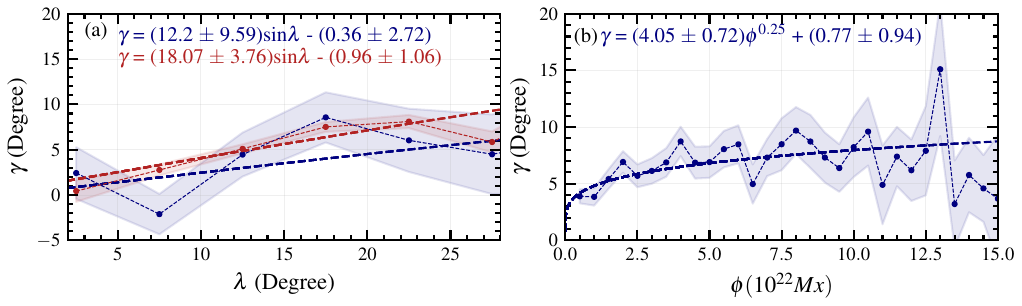}
\caption{(a) Displays Joy's law dependence for tracked BMRs in two different flux bins (1$\times$10$^{21}$~Mx -- 1$\times$10$^{22}$~Mx (blue) and 1$\times$10$^{22}$~Mx -- 1$\times$10$^{23}$~Mx (red)). The dashed line represents
the sinusoidal fit for both the flux bins with fitting parameters mentioned in the panel. (b) Mean  tilt (and the error by shaded region) in each flux bin as a function of the absolute flux.}
\label{fig:fig3}
\end{figure}

The initial inspection of the tracked data of BMRs shows the familiar tilt distribution and Joy's Law relation as shown by previous studies \citep[]{DS2010, WS1991} (also see Figure 4 in \citet{K2023}). To evaluate the flux dependence on tilt in the data, BMRs were segregated based on their assigned flux strengths into two different bins, 1$\times$10$^{21}$~Mx -- 1$\times$10$^{22}$~Mx (Bin-1) and 1$\times$10$^{22}$~Mx -- 1$\times$10$^{23}$~Mx (Bin-2). \Fig{fig:fig3}(a) shows Joy's law behavior in both the bins (Bin-1 in blue and Bin-2 in red) along with the sinusoidal fit. 
Notably, a greater degree of Joy's law scatter is evident for lower flux ranges, and the error bars of fitting parameters overlap, failing to show a conclusive tilt dependency on the measured flux of the BMR.  In \Fig{fig:fig3}(b), we further delve into flux dependency on tilt. Mean tilt values of BMRs in flux bins of 5 $\times$ 10$^{21}$~Mx are depicted against absolute total flux, accompanied by a fitted curve of the form $\gamma$ = a$\phi^{1/4}$ + b. However, a strong tilt dependence on flux is not yet apparent. A detailed study exploring this will be presented in our follow-up manuscript.

\begin{figure}[t]
\centering
\includegraphics[width=\textwidth]{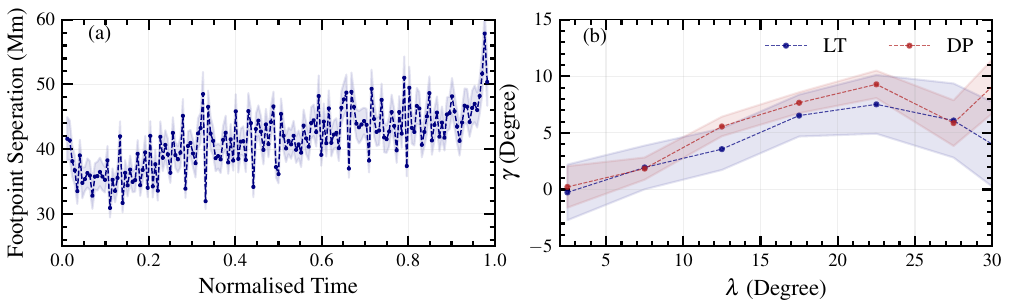}
\caption{(a) General evolution of foot point separation of BMR over its lifetime. (b) Joy’s law variation at the initial phase of the BMR. Blue represents Joy's law plot of the BMRs in LT class in the first 33\% of their lifetime. Red is the same as the blue curve but for BMRs in the DP class, which emerge in the nearside.}
\label{fig:fig4}
\end{figure}

We now proceed to evaluate the typical evolution of BMR foot points over BMR's lifetime. For this, we standardize the lifespan of LT class BMRs relative to the longest-living BMR within this class. Subsequently, we calculate the average foot point separation within each normalized lifespan interval to assess the overall progression of footprint separation, which is depicted in \Fig{fig:fig4}(a). The figure shows a consistent rise in foot point separation during the initial phase of BMR evolution, reaching saturation at approximately half of the BMR's lifespan. The observed sustained increase is in support of the thin-flux tube model.

Finally, in order to assess whether the BMRs exhibit definite tilt at an early stage of their evolution, we compile all BMRs within the LT class. We then examine the behavior of Joy's law during the initial phase of these BMRs. To do this, we compute the average values of tilt and emerging latitude within the initial 33\% of their lifespan and evaluate Joy's Law for this period, as illustrated in \Fig{fig:fig4}(b) in blue. As for DP class BMRs, we collect the BMRs that emerge on the near side, calculate the average values of tilt and emerging latitude over the entire tracking phase, and subsequently evaluate Joy's law (Figure 4(b) in red). The results indicate that during the initial phase of a BMR's lifespan, Joy's law is adhered to in both cases. This suggests that BMRs already exhibit a distinct tilt even in the early stages of AutoTAB tracking, which aligns with the findings of \citet{SK2008} and supports the thin-flux tube model.

\section{Conclusion}
A statistical study of the evolving magnetic and morphological properties of BMRs is crucial in the context of understanding solar magnetism. However, a comprehensive catalog of tracked data of BMRs is limited. To address this, we introduce an automatic algorithm to track the BMRs (AutoTAB). AutoTAB utilizes binary maps of identified bipolar regions and verifies their overlap in subsequent time frames for tracking purposes. AutoTAB identifies and tracks 9232 BMRs in the period of 1996-2022. AutoTAB only tracks the BMR for their evolution in the nearside of the Sun. Consequently, the dataset encompasses BMRs tracked throughout their entire lifespan and those tracked only during their evolutionary stages. 

The tracked BMRs were grouped into various classes based on their lifetime, namely, Short-lived (lives for less than 8 hr), Lifetime (emerges and disperses on the visible surface), and Disk passage (coming from and/or going to the far side of the Sun). The disk passage BMRs constitute the bigger BMRs with higher flux contents, and the short-lived group consists of mostly ephemeral regions. This dataset is further used to study the collective statistics of the BMRs. Our analysis reveals that the tracked BMRs exhibits behavior consistent with typical solar cycle patterns, with the emergence latitude of the BMRs at various times displaying butterfly wing-like patterns in latitude-time distribution plots. Furthermore, the tracked BMRs show familiar tilt distribution and tilt-latitude plots agreeing with previous studies. Further analyses were conducted to assess the validity of the thin-flux tube model for BMR formation. We observe that foot point separation of the BMRs of the LT class increases over the initial phase of their lifetime to saturate in the later phase, indicating the upward migration of magnetic field bundles from the convection zone towards the surface. Our analysis shows a well-defined Joy's law pattern in the initial phase of tracking, suggesting that BMRs emerge with a definite tilt. These findings support the thin-flux tube model. However, we do not find any strong and conclusive dependence of tilt on the flux of the BMR. These ideas will be further explored in the forthcoming article.

\section{Acknowledgment}
We thank IAU for the travel grant awarded to us, making it possible to attend the symposium in person in Armenia.

\bibliographystyle{iaulike}

\begin{thebibliography}{}

\bibitem[{Babcock}, 1961]{B1961}
{Babcock}, H.~W. 1961, {The Topology of the Sun's Magnetic Field and the 22-YEAR Cycle.}
\newblock {\em \apj}, 133, 572.

\bibitem[{Biswas} et~al., 2023]{BKK23}
{Biswas}, A., {Karak}, B.~B., \& {Kumar}, P. 2023, {Exploring the reliability of polar field rise rate as a precursor for an early prediction of solar cycle}.
\newblock {\em \mnras}, 526(3), 3994--4003.

\bibitem[{Dasi-Espuig} et~al., 2010]{DS2010}
{Dasi-Espuig}, M., {Solanki}, S.~K., {Krivova}, N.~A., {Cameron}, R., \& {Pe{\~n}uela}, T. 2010, {Sunspot group tilt angles and the strength of the solar cycle}.
\newblock {\em \aap}, 518, A7.

\bibitem[{D'Silva} and {Choudhuri}, 1993]{DC1993}
{D'Silva}, S. \& {Choudhuri}, A.~R. 1993, {A theoretical model for tilts of bipolar magnetic regions}.
\newblock {\em \aap}, 272, 621.

\bibitem[{Fan} et~al., 1994]{FF1994}
{Fan}, Y., {Fisher}, G.~H., \& {McClymont}, A.~N. 1994, {Dynamics of Emerging Active Region Flux Loops}.
\newblock {\em \apj}, 436, 907.

\bibitem[{Golubeva} et~al., 2023]{EB2023}
{Golubeva}, E.~M., {Biswas}, A., {Khlystova}, A.~I., {Kumar}, P., \& {Karak}, B.~B. 2023, {Probing the variations in the timing of the Sun's polar magnetic field reversals through observations and surface flux transport simulations}.
\newblock {\em \mnras}, 525(2), 1758--1768.

\bibitem[{Hale} et~al., 1919]{HE1919}
{Hale}, G.~E., {Ellerman}, F., {Nicholson}, S.~B., \& {Joy}, A.~H. 1919, {The Magnetic Polarity of Sun-Spots}.
\newblock {\em \apj}, 49, 153.

\bibitem[{Jha} et~al., 2020]{JK2020}
{Jha}, B.~K., {Karak}, B.~B., {Mandal}, S., \& {Banerjee}, D. 2020, {Magnetic Field Dependence of Bipolar Magnetic Region Tilts on the Sun: Indication of Tilt Quenching}.
\newblock {\em \apjl}, 889(1), L19.

\bibitem[{Jha} et~al., 2021]{JP2021}
{Jha}, B.~K., {Priyadarshi}, A., {Mandal}, S., {Chatterjee}, S., \& {Banerjee}, D. 2021, {Measurements of Solar Differential Rotation Using the Century Long Kodaikanal Sunspot Data}.
\newblock {\em \solphys}, 296(1), 25.

\bibitem[{Karak}, 2020]{Kar20}
{Karak}, B.~B. 2020, {Dynamo Saturation through the Latitudinal Variation of Bipolar Magnetic Regions in the Sun}.
\newblock {\em \apjl}, 901(2), L35.

\bibitem[{Karak}, 2023]{K2023}
{Karak}, B.~B. 2023, {Models for the long-term variations of solar activity}.
\newblock {\em Living Reviews in Solar Physics}, 20(1), 3.

\bibitem[{Karak} and {Brandenburg}, 2016]{KB2016}
{Karak}, B.~B. \& {Brandenburg}, A. 2016, {Is the Small-scale Magnetic Field Correlated with the Dynamo Cycle?}
\newblock {\em \apj}, 816(1), 28.

\bibitem[{Karak} and {Miesch}, 2017]{KM17}
{Karak}, B.~B. \& {Miesch}, M. 2017, {Solar Cycle Variability Induced by Tilt Angle Scatter in a Babcock-Leighton Solar Dynamo Model}.
\newblock {\em \apj}, 847, 69.

\bibitem[{Karak} and {Miesch}, 2018]{KM18}
{Karak}, B.~B. \& {Miesch}, M. 2018, {Recovery from Maunder-like Grand Minima in a Babcock--Leighton Solar Dynamo Model}.
\newblock {\em \apjl}, 860, L26.

\bibitem[{Kosovichev} and {Stenflo}, 2008]{SK2008}
{Kosovichev}, A.~G. \& {Stenflo}, J.~O. 2008, {Tilt of Emerging Bipolar Magnetic Regions on the Sun}.
\newblock {\em \apjl}, 688(2), L115.

\bibitem[{Leighton}, 1964]{L1964}
{Leighton}, R.~B. 1964, {Transport of Magnetic Fields on the Sun.}
\newblock {\em \apj}, 140, 1547.

\bibitem[{Mordvinov} et~al., 2022]{MK2022}
{Mordvinov}, A.~V., {Karak}, B.~B., {Banerjee}, D., {Golubeva}, E.~M., {Khlystova}, A.~I., {Zhukova}, A.~V., \& {Kumar}, P. 2022, {Evolution of the Sun's activity and the poleward transport of remnant magnetic flux in Cycles 21-24}.
\newblock {\em \mnras}, 510(1), 1331--1339.

\bibitem[{Mu{\~n}oz-Jaramillo} et~al., 2016]{JW2016}
{Mu{\~n}oz-Jaramillo}, A., {Werginz}, Z., {Vargas-Acosta}, J.~P., {DeLuca}, M., {Windmueller}, J.~C., {Zhang}, J., {Longcope}, D., {Lamb}, D., {DeForest}, C., {Vargas-Dom{\'\i}nguez}, S., {Harvey}, J., \& {Martens}, P.
\newblock {The best of both worlds: Using automatic detection and limited human supervision to create a homogenous magnetic catalog spanning four solar cycles}.
\newblock In {\em 2016 IEEE International Conference on Big Data (Big Data} 2016,, pp. 3194--3203.

\bibitem[{Parker}, 1955]{P1955}
{Parker}, E.~N. 1955, {The Formation of Sunspots from the Solar Toroidal Field.}
\newblock {\em \apj}, 121, 491.

\bibitem[{Scherrer} et~al., 1995]{SB1995}
{Scherrer}, P.~H., {Bogart}, R.~S., {Bush}, R.~I., {Hoeksema}, J.~T., {Kosovichev}, A.~G., {Schou}, J., {Rosenberg}, W., {Springer}, L., {Tarbell}, T.~D., {Title}, A., {Wolfson}, C.~J., {Zayer}, I., \& {MDI Engineering Team} 1995, {The Solar Oscillations Investigation - Michelson Doppler Imager}.
\newblock {\em \solphys}, 162, 129--188.

\bibitem[{Scherrer} et~al., 2012]{SS2012}
{Scherrer}, P.~H., {Schou}, J., {Bush}, R.~I., {Kosovichev}, A.~G., {Bogart}, R.~S., {Hoeksema}, J.~T., {Liu}, Y., {Duvall}, T.~L., {Zhao}, J., {Title}, A.~M., {Schrijver}, C.~J., {Tarbell}, T.~D., \& {Tomczyk}, S. 2012, {The Helioseismic and Magnetic Imager (HMI) Investigation for the Solar Dynamics Observatory (SDO)}.
\newblock {\em \solphys}, 275(1-2), 207--227.

\bibitem[{Sreedevi} et~al., 2023]{SJ2023}
{Sreedevi}, A., {Jha}, B.~K., {Karak}, B.~B., \& {Banerjee}, D. 2023, {AutoTAB: Automatic Tracking Algorithm for Bipolar Magnetic Regions}.
\newblock {\em \apjs}, 268(2), 58.

\bibitem[{Sreedevi} and {Jha}, 2023]{SJ2023Symp}
{Sreedevi}, A.~B. \& {Jha}, B.~K. 2023, {Automatic Algorithm for Tracking Bipolar Magnetic Regions}.
\newblock {\em IAU Symposium}, 372, 97--99.

\bibitem[{Stenflo} and {Kosovichev}, 2012]{SK2012}
{Stenflo}, J.~O. \& {Kosovichev}, A.~G. 2012, {Bipolar Magnetic Regions on the Sun: Global Analysis of the SOHO/MDI Data Set}.
\newblock {\em \apj}, 745(2), 129.

\bibitem[{Tlatov} and {Pevtsov}, 2014]{TP2014}
{Tlatov}, A.~G. \& {Pevtsov}, A.~A. 2014, {Bimodal Distribution of Magnetic Fields and Areas of Sunspots}.
\newblock {\em \solphys}, 289, 1143--1152.

\bibitem[{Wang} and {Sheeley}, 1991]{WS1991}
{Wang}, Y.~M. \& {Sheeley}, N.~R., J. 1991, {Magnetic Flux Transport and the Sun's Dipole Moment: New Twists to the Babcock-Leighton Model}.
\newblock {\em \apj}, 375, 761.

\end{thebibliography}

\end{document}